# Topological phenomena at topological defects

Zhi-Kang Lin[1], Qiang Wang[2], Yang Liu[1], Haoran Xue[2], Baile Zhang[2, †], Yidong Chong[2, †], and Jian-Hua Jiang[1, †]

[1]*Institute of Theoretical and Applied Physics, School of Physical Science and Technology & Collaborative Innovation Center of Suzhou Nano Science and Technology, Soochow University, 1 Shizi Street, Suzhou, 215006, China*

[2]*Division of Physics and Applied Physics, School of Physical and Mathematical Sciences, Nanyang Technological University, Singapore 637371, Singapore*

[†]Correspondence should be addressed to: blzhang@ntu.edu.sg (Baile Zhang), yidong@ntu.edu.sg (Yidong Chong), or jianhuajiang@suda.edu.cn (Jian-Hua Jiang).

## Abstract
There are two prominent applications of the mathematical concept of topology to the physics of materials: band topology, which classifies different topological insulators and semimetals, and topological defects that represent immutable deviations of a solid lattice from its ideal crystalline form. While these two classes of topological phenomena have generally been treated as separate topics, recent experimental advancements have begun to probe their intricate and surprising interactions, in real materials as well as synthetic metamaterials. Topological lattice defects in topological materials offer a platform to explore a diverse range of novel phenomena, such as topological pumping via topological defects, embedded topological phases, synthetic dimensions, and non-Hermitian skin effects. In this Perspective, we survey the developments in this rapidly moving field, and give an outlook of its impact on materials science and applications.

1. **Introduction**

Defects exist almost in every material. They appear due to imperfections in material growth processes, which are almost inevitable. Despite the word "defect" carrying negative connotations, they can play valuable roles in material applications, such as the regulation of semiconductors' electrical properties through the deliberate introduction of impurities. Defects can take on many forms, but certain types of crystallographic defects, called topological defects (TDs), are especially robust. TDs are characterized by the topology of the surrounding atomic lattice, making them highly resistant to elimination by any local lattice deformation [1, 2].

Although defects have long been studied in the context of conventional materials, recent works have shown that TDs can play exceptional roles in the so-called topological materials, such as topological insulators (TIs) [3-27] and topological semimetals [9, 28-30]. It is well known that topological materials are characterized by the topology of their band structures, as defined in momentum space, with nontrivial topologies giving rise to protected states at sample boundaries (surfaces, edges, and corners). However, topological modes can also occur on TDs in topological materials, arising from the interplay between the real-space topology associated with the TDs and the momentum-space topology underlying topological materials. Originally predicted and studied in theoretical works [3-30], these topological defect modes have only been successfully studied experimentally in the past few years [31-58]. These recent experimental achievements have taken place not only in condensed matter systems, but also on metamaterial platforms—in photonics, acoustics, and electrical circuits. In these materials, TDs have demonstrated their extraordinary roles in probing bulk topology

[40-45], inducing rich topological phenomena [31-36, 46-55], giving rise to cutting-edge applications [37-39], changing material properties [56-61], and synergy with other phenomena such as non-Hermitian effects and topological pumping [62-66].

There is a very wide variety of topological defect states, based on different combinations of TDs and topological materials. Some examples that are realized recently include Dirac vortex states [31-39], fractionally charged modes bound to disclinations in topological crystalline insulators (TCIs) [40, 41], helical one-dimensional (1D) modes bound to dislocations in weak topological insulators [42-44], and zero-dimensional (0D) topological modes at dislocations [48-50] and disclinations [51, 52] in TCIs. Moreover, topological defect states should not be regarded as conceptually straightforward generalizations of topological boundary states. In many cases, they allow access to novel physical phenomena, such as the ability to characterize topological features of TCI bands that are inaccessible to boundary probes [17-20, 40, 41], or probing the response of topological insulators to singular magnetic fluxes [6, 11]. Recent experimental studies have begun to explore many fresh possibilities, including topological states on finite spatial segments [46], vortex states induced by disclinations [47], TD-induced three-dimensional topological states of light [53], and spectral flows tied to the second Chern number [54]. Numerous other intriguing predictions, such as changing system topology by TDs [59], embedded topological phases [60, 61], topological disclination pumping [62], and non-Hermitian skin effects at TDs [63-66] remain to be verified in experiments.

In this perspective, we summarize recent progress on the new research frontier of TDs in topological materials. We survey the underlying physical principles and discuss how they are realized in a diverse set of material platforms. We also provide an outlook for this field, discussing challenges and possible future developments, as well as their impacts on materials science and applications. We hope this perspective will stimulate interest in a rapidly-evolving topic that has, to our knowledge, yet to be the subject of a comprehensive review. Due to space limitations, we omit discussions on the interesting role that TDs can also play in topological superconductors, superfluids, and strongly correlated systems.

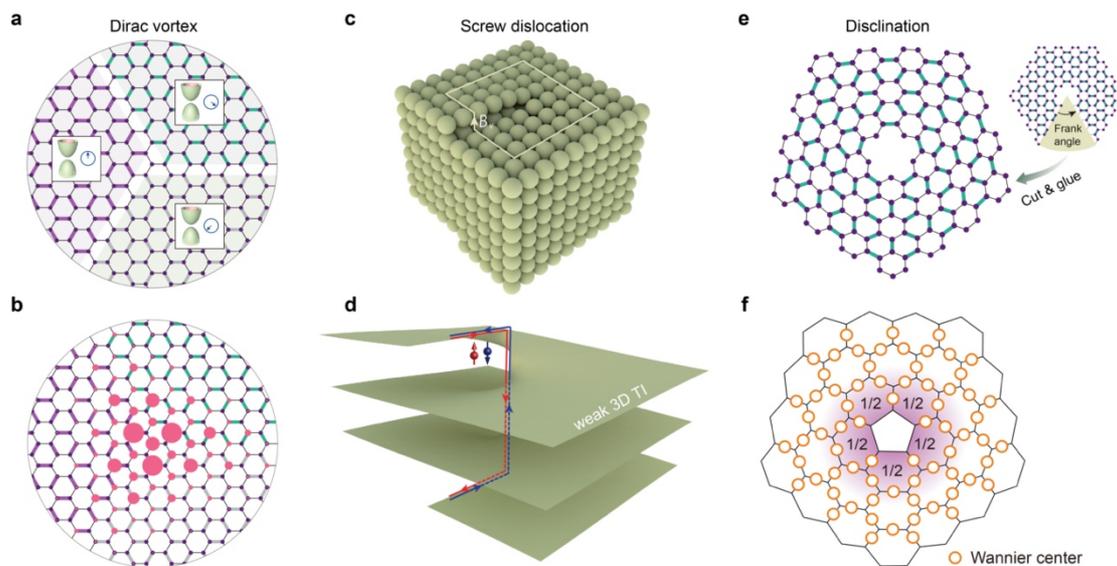

**Figure 1 | Topological defects and emergent phenomena. a**, A honeycomb lattice with Kekulé modulation. Each sector (spanning 120° in the illustration) has Dirac mass term with a different phase, forming a vortex called a Dirac vortex. **b**, A Majorana-like localized mode at the center of a Dirac vortex. The mode intensity is represented by the

red dots of different sizes. **c**, A screw dislocation in a cubic lattice. The Burgers vector is denoted as $B_v$. **d**, Helical 1D modes bound to a screw dislocation in a 3D weak TI. Spin-up and spin-down states are locked to the two opposite propagation directions, as indicated by the red and blue arrows. **e**, A disclination in a hexagonal topological lattice with dimerized nearest-neighbor couplings, formed by cutting off one sector (characterized by a Frank angle) and gluing the two cuts, as illustrated in the top-right inset. **f**, Fractional charge at a disclination, labelled by the numbers in the central five unit-cells.

## 2. Theoretical overview

In the past decades, physicists have discovered a cornucopia of topological phases characterized by various topological band invariants, such as TIs, Floquet TIs, TCIs (including HOTIs and fragile TIs), and topological semimetals [67]. Their most well-known feature is the bulk-boundary correspondence, which ties the existence of boundary modes to the nontrivial topology of the bulk band structure. For instance, Chern insulators exhibit chiral edge states that travel unidirectionally along their boundaries. Another property of topological phases, which has been much less studied, is that they have unique responses to TDs [3-30] including local modes (termed here as "TD modes") and fractional charges bound to TDs. These are governed by bulk-defect correspondences—a rich set of correspondence principles on equal footing with the bulk-boundary correspondence.

The simplest example of a TD mode is the kink mode in the Su-Schrieffer–Heeger (SSH) model, a topologically protected eigenstate bound to a 0D dislocation in the 1D lattice. In the continuum limit, the kink mode corresponds to the Jackiw-Rebbi solution for a 1D Dirac equation with a sign-switching defect in the mass field [3]. Since the bound state's energy is pinned to the middle of the band gap, in the condensed-matter context it carries 1/2 elementary charge. In two dimensions (2D), Jackiw and Rossi derived similar zero modes for the Dirac equation in a vortex background [4]. The zero mode is bound to the vortex core and is guaranteed by an index theorem tied to the vorticity. Such modes can be realized in honeycomb lattices with a vortex-like distortion called a Kekulé modulation [5] (Figs. 1**a**, 1**b**).

A theoretical framework for classifying TDs in TIs was devised by Teo and Kane based on the Altland-Zirnbauer "tenfold way" [6]. In this framework, the protected TD modes are determined by two pieces of information: the Altland-Zirnbauer class of the TI, and the difference in spatial dimensionality between the material and the TD. For example, the classification states that a screw dislocation in a three-dimensional (3D) weak TI (Fig. 1**c**) hosts helical 1D modes bound to the dislocation—a phenomenon that had first been discovered by Ran *et al.* [7]. The TD modes' existence is dictated jointly by the weak topological invariants of the valence bands and the Burgers vector of the TD, and is robust against disorders [7-9] (Fig. 1**d**). The lattice distortion induced by the dislocation induces an effective or "fictitious" gauge field in the long-wavelength limit, equivalent to that generated by a Dirac string running along the dislocation and carrying half a magnetic flux quantum [6, 17]. The binding of zero modes to Dirac strings is a signature feature of TIs [17, 18], and one that is thus seen to be intimately tied to the bulk-defect correspondence principle.

Another notable TD mode arises in a 2D honeycomb lattice with a 2D Chern insulator phase induced by time-reversal breaking. For instance, a disclination can be generated by a cut-and-glue construction [10], producing a central point defect where a hexagonal unit-cell is replaced by a pentagon (equivalent to placing the lattice on a conical surface with a singularity of spatial curvature at the apex) [11]. The resulting lattice distortion can be interpreted as a generalized Dirac string centered on the point defect, carrying an effective gauge field and binding a TD mode whose chirality is tied to the bulk Chern number.

TDs also play an important role in topological crystalline insulators (TCIs), a diverse family of topological phases protected by lattice symmetries, falling outside of the Altland-Zirnbauer scheme. For instance, a disclination in a TCI can disrupt the lattice rotational symmetry (Fig. 1**e**) and the charge configuration protected by such symmetry, yielding fractionally charged zero modes [12-16] (Fig. 1**f**). TD modes based on the bulk-disclination correspondence can provide more information about TCIs than the bulk-edge correspondence, enabling a resolution to the long-standing inadequacy of boundary-based probes for TCIs due to the reduction of lattice symmetry on the boundaries [40, 41]. In addition, TD modes can enable features that are inaccessible to edge or corner boundaries, such as "embedded topological phases" [60,61] and non-standard topological pumping via TDs [62].

Turning now to topological semimetals, the bulk-dislocation correspondence in these materials manifests at the band degeneracy energy, with TD modes co-existing with the bulk zero modes under external magnetic fields [9]. Moreover, it is possible in topological semimetals to use TD-induced effective gauge fields to access the chiral anomaly [28, 29]. These and other TD-related phenomena can be used to probe the topological charge of the band degeneracies in topological semimetals [30].

The many possible configurations of TDs in various materials provide a rich set of possibilities for accessing diverse topological phenomena. Moreover, the incorporation of synthetic dimensions and dynamic modulations provides the opportunity to explore TDs in higher dimensional topological lattices, which can express different behaviors [22-24, 53-55]. TD modes may also have applications in various technology frontiers. For instance, in metamaterials (see Section 3), 1D TD modes with spin-momentum locking can enable new forms of wave-guiding for classical waves [43, 44], and Dirac vortices can be used to realize braiding of photonic states [35] and high-performance topological surface lasers [38, 39].

| Topological Phases | | Dimension and Type of Topological Defects | Key Features | Experimental Platforms |
|---|---|---|---|---|
| Topological insulators | 2D Chern insulator | 0D conical singularity [10, 11] | Gravitational anomaly | Optical resonators [56, 57] |
| | | 0D dislocation [17] | 0D bound modes | Photonic crystals [50] |
| | 3D Chern insulator | 1D helix modulation [27] | Chiral states | |
| | 2nd Chern insulator | 2D dislocation | Spectral flows | Photonic + synthetic dimensions [54] |
| | 2D weak TI | 0D dislocation | 0D bound modes | Mechanical [48,49] |
| | 3D weak TI | 1D dislocation [7-9] | Helical 1D modes | Electronic [42], acoustic [43, 44], photonic [53] |

| Topological semimetal | Weyl semimetal | 1D dislocation/disclination [29, 30] | Vortex states; defect-induced chiral anomaly | Acoustic [47] |
|---|---|---|---|---|
| | Dirac semimetal | 1D dislocation [28] | Defect-induced chiral anomaly | |
| Topological crystalline insulator | 2D/3D HOTI | 0D disclination [15]; 0D and 1D dislocations [20]; 2D partial lattice defect [21] | Fractional charges; Wannier cycles; bound states | Photonic [40]; transmission line [41]; acoustic [45,51]; electrical circuits [58] |
| Other | Floquet TI | 1D dislocations [22, 23]; 0D phase modulation | Helical 1D states; defect mode resonance | Photonic [55] |
| | Valley-Hall insulator | Disclinations; disclination pair induced internal edge | Bound states; internal edge mode | Acoustic [52]; photonic [46] |
| | Kekulé graphene | 0D supercell modulation [5, 25, 26] | Dirac-vortex modes | Acoustic [31, 37]; mechanical [32,33]; photonic [34-36, 38, 39] |

**Table I | Prominent TD-induced phenomena in topological materials.** Various topological phases are characterized by notable TDs and TD-induced phenomena, along with recent experimental realizations in various material platforms.

## 3. Platforms

TDs are quite common in solids and are especially abundant in nanomaterials [68]. For instance, TDs commonly exist at grain boundaries where lattice matching fails [69, 70] (see Section 7). Although TDs are hard to control in these materials, they can still be affected by the growth conditions [68]. The TD modes can be probed experimentally via scanning tunneling microscopes and transmission electron microscopes.

In metamaterials, TDs can be introduced, fabricated, and tailored with exceptional ease, and these platforms have shown advantages in recent studies [31-41, 43-58]. These synthetic lattices, in which classical waves (light, sound, etc.) take on the role of the electron wavefunction, can exhibit many of the features of real materials. Examples include acoustic and photonic metamaterials, which have different merits and shortcomings in sample fabrication and measurements. In particular, in 3D systems, acoustic waves are spinless, while photons have spin degrees of freedom. Electrical circuits [41, 58], coupled waveguide arrays [34, 35] and lattices of coupled optical fibers [53] also fall into this category.

Aside from the ease of fabrication, metamaterials possess several other advantages. Depending on the setup, they can also be probed with excellent position, wavevector, and frequency resolution at room temperature [31-41]. Dynamic tuning of TDs in metamaterials may enable emergent phenomena such as TD pumping [62]. Finally, metamaterials provide access to effects such as Floquet driving [22-24, 53, 55], non-Hermiticity [63-66] and synthetic dimensions [24, 53, 54] which introduce further richness to the interplay between TDs and band topology.

## 4. Defects in TIs and topological semimetals

Dislocation modes in 2D Chern insulators were first experimentally realized and studied in a photonic crystal containing gyromagnetic rods (which break time-reversal symmetry) [50] (Fig. 2**a**). The dislocation, a 0D TD, was generated by removing half a column of unit-cells from a rectangular-lattice photonic crystal, and the TD mode was imaged by direct field measurements (Fig. 2**b**). In a non-lattice context, Schine *et al.* [56, 57] used a "twisted resonator" formed by carefully arranged mirrors to implement an optical quantum Hall phase defined on a synthetic space with a conical singularity (equivalent to a disclination in a lattice; see Section 2) and observed local modes at the conical singularity (Figs. 2**c**, 2**d**).

For 3D TIs, the dislocation modes predicted by Ran *et al.* [7] (see Section 2) have been observed in both condensed matter [42] and acoustic metamaterials [43]. For the former, Nayak *et al.* developed bismuth (Bi) crystals with well-isolated screw dislocations by cleaving from GdPtBi and used differential conductance measurements to observe the TD modes (Fig. 2**e**). In acoustics, dislocation modes were observed by Xue *et al.* in a weak acoustic 3D TI (Fig. 2**f**), implemented in a lattice of weakly-coupled resonators through the use of a pseudospin transformation that maps all inter-site hoppings to real values [43]. A different acoustic design, based on stacked layers of coupled ring resonators, has also been used by Ye *et al.* to observe dislocation modes [44]. Recently, with synthetic dimensions enabled by modulating coupled optical waveguides, such 1D wave-guiding has also been realized in a photonic waveguide array [53] (Fig. 2**g**).

In synthetic structures, it is even possible to realize four- or higher-dimensional lattices, allowing for the study of high-dimensional topological phases and TD modes that cannot possibly exist in real materials. For instance, using a 2D dielectric lattice with an additional two synthetic dimensions induced by parametric variations, Chen *et al.* have implemented a 4D second Chern insulator supporting gapless 2D dislocation modes [54], enabling a new understanding of trapped modes induced by edge dislocations in 2D.

There has also been notable research progress in TD-induced phenomena in topological semimetal phases. Wang *et al.* have developed an acoustic 3D Weyl lattice containing a 1D disclination that serves as an acoustic waveguide [47] (Fig. 2**h**). The structure is periodic along the *z* axis, and for wavenumber $k_z$ spanning the two Weyl points, the system maps into a 2D Chern insulator with a disclination (Fig. 2**i**). Moreover, the disclination-induced waveguide modes carry orbital angular momentum locked to the direction of propagation along the disclination line. TDs in topological semimetals have been theoretically predicted to give rise to a range of other intriguing phenomena [9, 28-30] (see Section 2) that have yet to be confirmed in experiments.

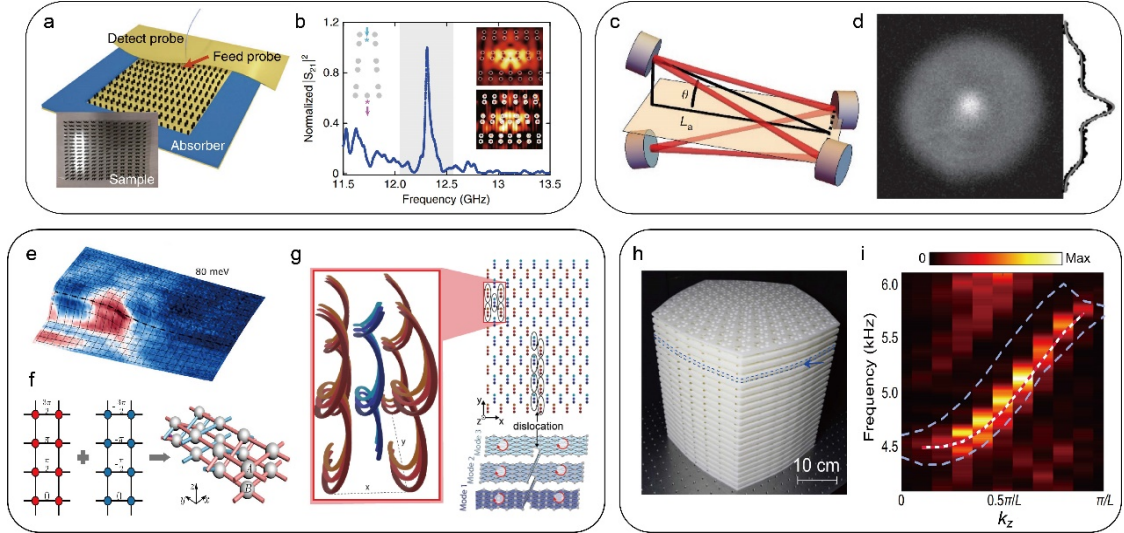

**Figure 2 | Topological defects in topological insulators and Floquet topological insulators. a-b,** Topological light-trapping on a dislocation. The experimental system consists of a 2D topological photonic crystal with a dislocation and a pump-probe setup, as shown in (a). The transmission across the dislocation as well as the simulated (upper) and measured (lower) field profiles of the topological bound state are plotted in (b). **c-d,** Conical singularity induced gravitational anomaly in photonic quantum Hall systems. **c,** The non-planar resonator, consisting of four mirrors. **d,** The measured local state density near the cone. **e,** Bismuth with a screw dislocation (color map shows the local density of electronic states). **f,** A 3D acoustic weak TI based on a pseudospin transformation, which supports pseudospin resolved helical 1D transport along a screw dislocation. **g,** Screw dislocations in a photonic Floquet TI with a synthetic dimension. **h-i**, Acoustic vortex states in an acoustic Weyl semimetal. **h,** A photo of the acoustic Weyl crystal with a disclination line at the center. **i,** The measured acoustic dispersion showing 1D localized states along the disclination axis. Images in **a** and **b** adapted from Ref. [50], Nature Publishing Group. Images in **c** and **d** adapted from Ref. [56], Nature Publishing Group. Image in **e** adapted from Ref. [42], American Association for the Advancement of Science. Image in **f** adapted from Ref. [43], American Physical Society. Image in **g** adapted from Ref. [53]. Images in **h** and **i** adapted from Ref. [47], Nature Publishing Group.

## 5. Defects in TCIs

The experimental study of TDs in TCIs has attracted a great deal of recent attention, with a particular focus on fragile and higher-order TIs. As TDs offer a diverse set of internal "boundaries", they can generate a rich set of novel effects in TCIs. Thanks to the development of metamaterials, many of these phenomena can be accessed relatively easily in acoustic and photonic experiments. For instance, TD modes have been experimentally observed in mechanical TCIs [49] and electrical circuits implementing multipole TIs [58]. Moreover, TD modes can also serve as versatile probes for the bulk topological characteristics of HOTIs and other TCI phases [17-20]. The response of a TCI to a disclination can involve the remarkable phenomenon of fractionally charged TD modes [18, 19, 40, 41] (Figs. 3**a**, 3**b**). Here, the bulk-disclination correspondence relates the fractional bound charge to the topological invariants of the TCI and the Frank angle of the disclination. In the experiments, which are based on photonic rather than electronic TCIs, the fractionalization was quantitatively probed using local photonic density-of-states measurements over a broad frequency range, interpreted via an analogy with the filling of electronic valence bands [40].

Although existing experimental demonstrations of disclination-induced fractional TD modes have been based on 2D systems [40, 41], similar phenomena may occur in higher dimensions (possibly including synthetic dimensions [24, 53, 54]). The exceptionally rich set of TCIs in 3D systems (with 230 space groups) may give rise to an abundance of different fractional TD modes which are yet to be explored.

TCIs support another distinct mechanism for the formation of TD modes, called topological Wannier cycles [45] (Fig. 3**c**). Topological Wannier cycles occur in spinless TCIs possessing well-defined Wannier orbitals, which are HOTIs supporting corner states (such materials are also called obstructed atomic insulators, since the Wannier centers lie away from the unit-cell center). In the absence of chiral symmetry, these corner states are ordinarily unprotected from merging into the bulk or edge continuum, but topological Wannier cycles provide a way to stabilize them. The underlying mechanism originates from the effect of a gauge flux threading the Wannier orbitals at the same location but with different spatial symmetries; when a gauge flux is threaded, the Wannier orbitals must cyclically transform into one another to fulfill simultaneously the space symmetry and gauge invariance, resulting in a cyclic spectral flow among the bulk bands [71]. This phenomenon has been quantitatively demonstrated using an acoustic structure with a screw dislocation [45].

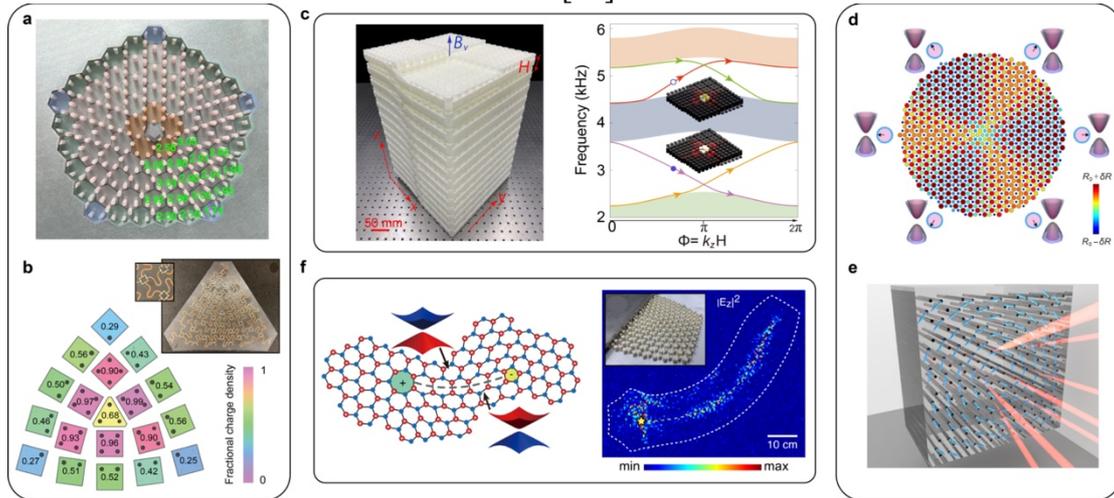

**Figure 3 | Topological defects in topological crystalline insulators. a**, **b**, Disclination-induced fractional charges in topological crystalline insulators. The numbers in each plot indicate the measured mode number in the corresponding unit-cell. **c**, Screw dislocation induced topological Wannier cycles in an acoustic HOTI. The left subplot shows the picture of the sample with a screw dislocation with the Burgers vector denoted by $B_v$ and the blue arrow. The right subplot shows the topological Wannier cycle spectrum and the profiles of the dislocation modes. **d**, An acoustic Dirac vortex cavity realized by modulating the radii of the rigid cylinders in the unit-cells. **e**, A photonic Dirac vortex cavity composed of laser-written waveguide arrays with a non-uniform Kekulé-distortion. **f**, Internal edge states induced by a disclination pair in a photonic valley Hall system. The left subplot illustrates the lattice geometry and the flip of the mass of the Dirac cone across the disclination line. The right subplot shows the measured intensity distribution of the electric field, excited by a point source placed at the position denoted by the yellow star. Image in **a** reproduced from Ref. [40], Nature Publishing Group. Image in **b** reproduced from Ref. [41], Nature Publishing Group. Image in **b** reproduced from Ref. [41], Nature Publishing Group. Image in **c** adapted from Ref. [45], Nature Publishing Group. Image in **d** adapted from Ref. [31], American

Physical Society. Image in **e** adapted from Ref. [34], American Physical Society. Image in **f** adapted from Ref. [46], American Physical Society.

**6. Other types of topological defects in topological materials**

As discussed in Section 2, the Dirac vortex is an elementary TD that can bind a single mid-gap state called the Jackiw-Rossi zero mode. It was suggested by Hou *et al.* [5] that the Dirac vortex can be realized via a Kekulé-textured honeycomb lattice, in which the contents of each unit-cell are subjected to a modulation parameterized by a phase that varies with the azimuthal angle around a given position (the vortex core). This modulation, which breaks the translational symmetry of the lattice, manifests as a complex-valued Dirac mass field that undergoes a phase winding when encircling the vortex core in real space (Fig. 1**a**). The Kekulé-distortion leads to the formation of the topological zero mode localized at the vortex core, and residing only on one sublattice (Fig. 1**b**).

Recently, the Dirac vortex and its associated zero mode have been realized and studied in a rapid succession of experiments [31-39]. The first experimental realization used acoustic structures, with the Kekulé texture created by varying the radii of rigid cylinders [31] (Fig. 3**d**). Similar zero modes were subsequently observed in mechanical [32, 33] and photonic [34, 35] systems. A photonic realization of the Dirac vortex in waveguide systems is illustrated in Fig. 3**e**, with the Kekulé texture created by displacing the waveguide centers from their default positions [34]. This platform has also been used to demonstrate a "braiding operation", a geometrically nontrivial process implemented by interchanging two different zero modes [35]. An analogous realization based on a silica photonic crystal fiber was proposed recently as a means of achieving robust transmission in optical fibers [25]. Exciting progress has also been achieved in applying the photonic Dirac vortex as a topological cavity for a vertical-cavity surface-emitting laser [38, 39].

Most of the phenomena described above are induced by an individual TD. However, there are also novel topological phenomena associated with the presence of multiple TDs. For example, Wang *et al.* [46] demonstrated that introducing a pair of disclinations (a pentagon and a heptagon) into a photonic valley Hall lattice generates an internal boundary extending between the two TDs, hosting robust valley Hall-like edge states (Fig. 3**f**). Such edge states can be used to achieve curved and free-form topological wave-guiding that is robust against disorder-induced localization. Similar disclination-based waveguiding has also recently been numerically studied in a mechanical (phononic) lattice [72].

**7. Outlook**

Topological defects provide a rich set of opportunities to generate novel behaviors in topological materials, beyond what is possible using ordinary sample boundaries and domain walls. Aside from the phenomena surveyed in the previous sections, others that have recently been explored include anomalous chiral modes induced by TDs in 3D time-reversal breaking Floquet systems [22], and the "embedding" of lower-dimensional topological phases in higher-dimensional topologically nontrivial materials, which can be observed by introducing TDs into the host lattices [60, 61].

Evidently, much of the recent progress in the experimental study of topological defect modes has been based on metamaterial platforms, with few exceptions [42]. Many real topological materials contain TDs, including well-known 2D materials, nanostructures, and nanomaterials (Figs. 4**a**-4**e**) [68]. Thus far, their physical consequences have been the subject of relatively few theoretical studies [73-79]. However, with the progress of technology (including improvements in scanning tunneling microscopy, atomic force microscopy, and transmission electron microscopy), as well as improvements in our theoretical understanding, TDs in topological materials are a promising subject of exploration. The rich geometries of TDs in various nanomaterials provide the opportunity to manipulate the motion of electrons and phonons in ways beyond what is possible with standard materials, boundaries, and domain walls, with implications for unconventional transport, optical, electronic, chemical, and phononic properties.

On the metamaterials front, we anticipate that TDs will increasingly be utilized to create new forms of waveguides, resonators, and other devices for controlling photonic, phononic and electrical waves. These synthetic lattices also hold considerable promise for exploring the interplay between TDs, band topology, and additional physical features such as non-Hermiticity, nonlinearity, and quantum effects. For instance, recently predicted intriguing non-Hermitian "skin effects" induced by TDs (Fig. 4**f**) can be realized and studied in these platforms [63-66].

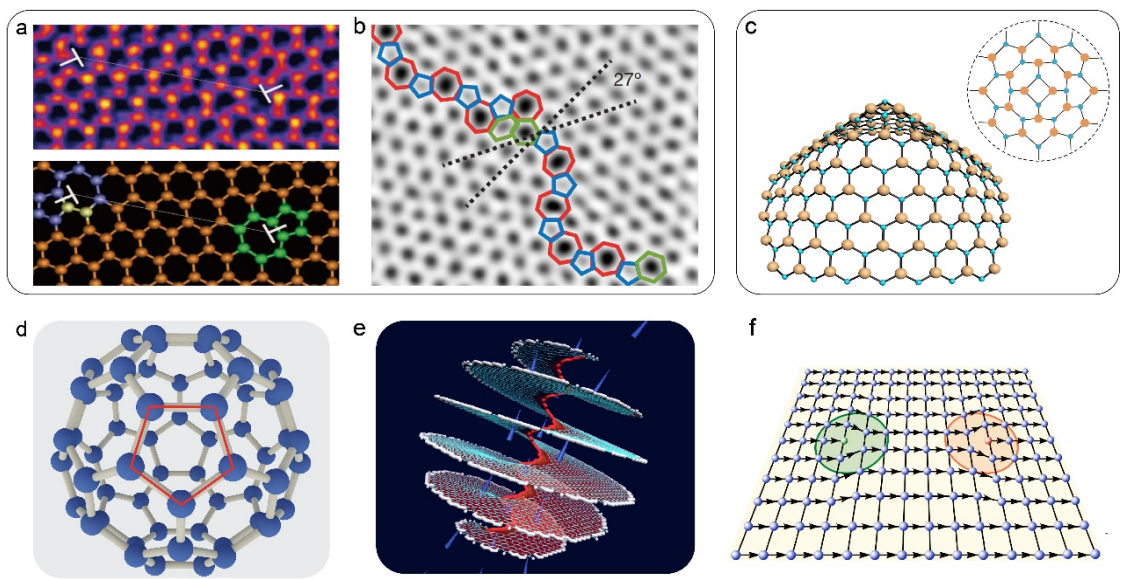

**Figure 4 | Topological defects in nanomaterials and non-Hermitian systems. a**, Scanning tunneling microscope image (top) and atomic model (bottom) of disclination pairs in graphene. The white markers denote the positions of the disclinations. **b**, A chain of disclinations at the grain boundary of graphene. **c**, Disclination at the tip of a boron nitride nanocone. Inset shows the top view of the nanocone around the tip, indicating the disclination structure. **d**, Disclination in a fullerene structure as indicated by the pink pentagon. **e**, Multi-layer graphene with a screw rotated Riemann surface. **f**, Non-Hermitian skin effects at a pair of dislocations indicated by the green and red circles. Image in **a** adapted from Ref. [73], American Association for the Advancement of Science. Image in **b** adapted from Ref. [70], Nature Publishing Group. Image in **c** adapted from Ref. [52]. Image in **e** reproduced from Ref. [75], American Chemical Society. Image in **f** adapted from Ref. [65], American Physical Society.

References:
1. Mermin, N. D. The topological theory of defects in ordered media. Rev. Mod. Phys. 51, 591-648 (1979).

2. Kleman, M. & Friedel, J. Disclinations, dislocations, and continuous defects: A reappraisal. *Rev. Mod. Phys.* **80**, 61-115 (2008).
3. Jackiw, R. & Rebbi, C. Solitons with fermion number 1/2. *Phys. Rev. D* **13**, 3398–3409 (1976).
4. Jackiw, R. & Rossi, P. Zero modes of the vortex-fermion system. *Nucl. Phys. B* **190**, 681–691 (1981).
5. Hou, C.-Y., Chamon, C. & Mudry, C. Electron fractionalization in two-dimensional graphene-like structures. *Phys. Rev. Lett.* **98**, 186809 (2007).
6. Teo, J. C. Y. & Kane, C. L. Topological defects and gapless modes in insulators and superconductors. *Phys. Rev. B* **82**, 115120 (2010).
7. Ran, Y., Zhang, Y. & Vishwanath, A. One-dimensional topologically protected modes in topological insulators with lattice dislocations. *Nat. Phys.* **5**, 298–303 (2009).
8. Imura, K.-I., Takane, Y. & Tanaka, A. Weak topological insulator with protected gapless helical states. *Phys. Rev. B* **84**, 035443 (2011).
9. de Juan, F., Rüegg, A. & Lee, D.-H. Bulk-defect correspondence in particle-hole symmetric insulators and semimetals. *Phys. Rev. B* **89**, 161117(R) (2014).
10. Rüegg, A. & Lin, C. Bound States of Conical Singularities in Graphene-Based Topological Insulators. *Phys. Rev. Lett.* **100**, 046401 (2013).
11. Biswas, R. R. & Son, D. T. Fractional charge and inter-Landau–level states at points of singular curvature. *Proc. Natl. Acad. Sci.* **113**, 8636-8641 (2016).
12. Teo, J. C. Y. & Hughes, T. L. Existence of Majorana fermion bound states on disclinations and the classification of topological crystalline superconductors in two dimensions. *Phys. Rev. Lett.* **111**, 047006 (2013).
13. Benalcazar, W. A., Teo, J. C. Y. & Hughes, T. L. Classification of two-dimensional topological crystalline superconductors and Majorana bound states at disclinations. *Phys. Rev. B* **89**, 224503 (2014).
14. Teo, J. C. Y. & Hughes, T. L. Topological defects in symmetry-protected topological phases. *Annu. Rev. Conden. Matter Phys.* 8, 211-237 (2017).
15. Li, T., Zhu, P., Benalcazar, W. A. & Hughes, T. L. Fractional disclination charge in two-dimensional $C_n$-symmetric topological crystalline insulators. *Phys. Rev. B* **101**, 115115 (2020).
16. Geier, M., Fulga, I. C. & Lau, A. Bulk-boundary-defect correspondence at disclinations in rotation-symmetric topological insulators and superconductors. *SciPost Phys.* **10**, 092 (2021).
17. Juričić, V., Mesaros, A., Slager, R.-J. & Zaanen, J. Universal probes of two-dimensional topological insulators: dislocation and π-flux. *Phys. Rev. Lett.* **108**, 106403 (2012).
18. Slager, R.-J., Mesaros, A., Juričić, V. & Zaanen, J. Interplay between electronic topology and crystal symmetry: Dislocation-line modes in topological band insulators. *Phys. Rev. B* **90**, 241403(R) (2014).
19. van Miert, G. & Ortix, C. Dislocation charges reveal two-dimensional topological crystalline invariants. *Phys. Rev. B* **97**, 201111(R) (2018).
20. Roy, B. & Juričić, V. Dislocation as a bulk probe of higher-order topological insulators. *Phys. Rev. Res.* **3**, 033107 (2021).
21. Queiroz, R., Fulga, I. C., Avraham, N, Beidenkopf, H. & Cano, J. Partial lattice defects in higher-order topological insulators. *Phys. Rev. Lett.* **123**, 266802 (2019).
22. Bi, R., Yan, Z., Lu, L. & Wang, Z. Topological defects in Floquet systems: Anomalous chiral modes and topological invariant. *Phys. Rev. B* **95**, 161115(R) (2017).

23. Nag, T. & Roy, B. Anomalous and normal dislocation modes in Floquet topological insulators. *Commun. Phys.* **4**, 157 (2021).
24. Lin, Q., Sun, X.-Q., Xiao, M., Zhang, S.-C. & Fan, S. A three-dimensional photonic topological insulator using a two-dimensional ring resonator lattice with a synthetic frequency dimension. *Sci. Adv.* **4**, eaat2774 (2018).
25. Lin, H. & Lu, L. Dirac-vortex topological photonic crystal fibre. *Light Sci. Appl.* **9**, 202 (2020).
26. Košata, J. & Zilberberg, O. Second-order topological modes in two-dimensional continuous media. *Phys. Rev. Res.* **3**, L032029 (2021).
27. Lu, L., Gao, H. & Wang, Z. Topological one-way fiber of second Chern number. Nat. Commun. 9, 5384 (2018).
28. Chernodub, M. N. & Zubkov, M. A. Chiral anomaly in Dirac semimetals due to dislocations. *Phys. Rev. B* **95**, 115410 (2017).
29. Sumiyoshi, H. & Fujimoto, S. Torsional chiral magnetic effect in a Weyl semimetal with a topological defect. *Phys. Rev. Lett.* **116**, 166601 (2016).
30. Soto-Garrido, R., Muñoz, E. & Juričić, V. Dislocation defect as a bulk probe of monopole charge of multi-Weyl semimetals. *Phys. Rev. Res.* **2**, 012043(R) (2020).
31. Gao, P., Torrent, D., Cervera, F., San-Jose, P., Sánchez-Dehesa, J. & Christensen, J. Majorana-like zero modes in Kekulé distorted sonic lattices. *Phys. Rev. Lett.* **123**, 196601 (2019).
32. Chen, C.-W. *et. al.* Mechanical analogue of a Majorana bound state. *Adv. Mater.* **31**, 1904386 (2019).
33. Ma, J., Xi, X., Li, Y. & Sun, X. Nanomechanical topological insulators with an auxiliary orbital degree of freedom. *Nat. Nanotechnol*. **16**, 576-583 (2021).
34. Menssen, A. J. *et. al.* Photonic topological mode bound to a vortex. *Phys. Rev. Lett.* **125**, 117401 (2020).
35. Noh, J. *et. al.* Braiding photonic topological zero modes. *Nat. Phys.* **16**, 989-993 (2020).
36. Gao, X. *et. al.* Dirac-vortex topological cavities. *Nat. Nanotechnol.* **15**, 1012-1018 (2020).
37. Xi, X., Ma, J. & Sun, X. A topological Dirac-vortex parametric phonon laser. Preprint at https://arxiv.org/abs/2107.11162
38. Yang, L., Li, G., Gao, X. & Lu, L. Topological-cavity surface-emitting laser. *Nat. Photon.* **16**, 279–283 (2022).
39. Ma, J. *et. al.* Room-temperature continuous-wave Dirac-vortex topological lasers on silicon. Preprint at https://arxiv.org/abs/2106.13838.
40. Liu, Y. *et. al.* Bulk-disclination correspondence in topological crystalline insulators. *Nature* **589**, 381-385 (2021).
41. Peterson, C. W. *et. al.* Trapped fractional charges at bulk defects in topological insulators. *Nature* **589**, 376-380 (2021).
42. Nayak, A. K. *et. al.* Resolving the topological classification of bismuth with topological defects. *Sci. Adv.* **5**, eaax6996 (2019).
43. Xue, H. *et. al.* Observation of dislocation-induced topological modes in a three-dimensional acoustic topological insulator. *Phys. Rev. Lett.* **127**, 214301 (2021).
44. Ye, L., Qiu, C., Xiao, M., Li, T., Du, J., Ke, M. & Liu, Z. Topological dislocation modes in three-dimensional acoustic topological insulators. *Nat. Commun.* **13**, 508 (2022).
45. Lin, Z.-K. *et. al.* Topological Wannier cycles induced by sub-unit-cell artificial gauge flux in a sonic crystal. *Nat. Mater.* **21**, 430-437 (2022).

46. Wang, Q., Xue, H., Zhang, B. & Chong, Y. D. Observation of protected photonic edge states induced by real-space topological lattice defects. *Phys. Rev. Lett.* **124**, 243602 (2020).
47. Wang, Q. *et. al.* Vortex states in an acoustic Weyl crystal with a topological lattice defect. *Nat. Commun.* **12**, 3654 (2021).
48. Paulose, J., Chen, B. G. & Vitelli, V. Topological modes bound to dislocations in mechanical metamaterials. *Nat. Phys.* **11**, 153–156 (2015).
49. Grinberg, I. H., Lin, M., Benalcazar, W. A., Hughes, T. L. & Bahl, G. Trapped state at a dislocation in a weak magnetomechanical topological insulator. *Phys. Rev. Appl.* **14**, 064042 (2020).
50. Li, F.-F. *et. al.* Topological light-trapping on a dislocation. *Nat. Commun.* **9**, 2462 (2018).
51. Deng, Y. *et. al.* Observation of degenerate zero-energy topological states at disclinations in an acoustic lattice. *Phys. Rev. Lett.* **128**, 174301 (2022).
52. Chen, Y. *et. al.* Observation of topological disclination states in non-Euclidean geometries. Preprint at https://arxiv.org/abs/2201.10039 (2022).
53. Lustig, E. *et. al.* Three-dimensional photonic topological insulator induced by lattice dislocations. Preprint at https://arxiv.org/abs/2204.13762 (2022).
54. Chen, X. D. *et. al.* Second Chern crystals in four-dimensional synthetic translation space with inherently nontrivial topology. Preprint at https://arxiv.org/abs/2112.05356 (2021).
55. Afzal, S. & Van, V. Trapping light in a Floquet topological photonic insulator by Floquet defect mode resonance. *APL Photon.* **6**, 116101 (2021).
56. Schine, N. *et. al.* Synthetic Landau levels for photons. *Nature* **534**, 671-675 (2016).
57. Schine, N. *et. al.* Electromagnetic and gravitational responses of photonic Landau levels. *Nature* **565**, 173-179 (2019).
58. Yamada, S. S. *et. al.* Bound states at partial dislocation defects in multipole higher-order topological insulators. *Nat. Commun.* **13**, 2035 (2022).
59. Barkeshli, M. & Qi, X.-L. Topological nematic states and non-Abelian lattice dislocations. *Phys. Rev. X* **2**, 031013 (2012).
60. Tuegel, T. I., Chua, V. & Hughes, T. L. Embedded topological insulators. *Phys. Rev. B* **100**, 115126 (2019).
61. Velury, S. & Hughes, T. L. Embedded topological semimetals. *Phys. Rev. B* **105**, 184105 (2022).
62. Xie, B.-Y., You, O. & Zhang, S. Topological disclination pump. Preprint at https://arxiv.org/abs/2104.02852 (2021).
63. Sun, X.-Q., Zhu, P. & Hughes, T. L. Geometric response and disclination induced skin effect in non-Hermitian systems. *Phys. Rev. Lett.* **127**, 066401 (2021).
64. Schindler, F. & Prem, A. Dislocation non-Hermitian skin effect. *Phys. Rev. B* **104**, L161106 (2021).
65. Bhargava, B. A., Fulga, I. C., Brink, J. V. D. & Moghaddam, A. G. Non-Hermitian skin effect of dislocations and its topological origin. *Phys. Rev. B* **104**, L241402 (2021).
66. Panigrahi, A., Moessner, R. & Roy, B. Non-Hermitian dislocation modes: Stability and melting across exceptional points. Preprint at https://arxiv.org/abs/2105.05244 (2021).
67. Wieder, B. J. *et. al.* Topological materials discovery from crystal symmetry. *Nat. Rev. Mater.* **7**, 196-216 (2022).
68. Gupta, S. & Saxena, A. A topological twist on materials science. *MRS Bull.* **39**, 265–279 (2014).

69. Yazyev, O. V. & Louie, S. G. Topological defects in graphene: Dislocations and grain boundaries. *Phys. Rev. B* **81**, 195420 (2010).
70. Huang, P. Y. *et. al.* Grains and grain boundaries in single-layer graphene atomic patchwork quilts. *Nature* **469**, 389-392 (2011).
71. Song, Z.-D., Elcoro, L. & Bernevig, B. A. Twisted bulk-boundary correspondence of fragile topology. *Science* **367**, 794–797 (2020).
72. Xia, B., Tong, L., Zhang, J., Zheng, S. & Man, X. Topological defect states in elastic phononic plates. Preprint at https://arxiv.org/abs/2102.07552 (2021).
73. Warner, J. H. *et. al.* Dislocation-driven deformations in graphene. *Science* **337**, 209-212 (2012).
74. Butz, B. *et. al.* Dislocations in bilayer graphene. *Nature* **505**, 533-537 (2014).
75. Xu, F., Yu, H., Sadrzadeh, A. & Yakobson, B. I. Riemann surfaces of carbon as graphene nanosolenoids. *Nano Lett.* **16**, 34 (2016).
76. Zhi, C., Bando, Y., Tang, C. & Golberg, D. Electronic structure of boron nitride cone-shaped nanostructures. *Phys. Rev. B* **72**, 245419 (2005).
77. Rüegg, A., Coh, S. & Moore, J. E. Corner states of topological fullerenes. *Phys. Rev. B* **88**, 155127 (2013).
78. Ochoa, H., Zarzuela, R. & Tserkovnyak, Y. Emergent gauge fields from curvature in single layers of transition-metal dichalcogenides. *Phys. Rev. Lett.* **118**, 026801 (2017).
79. Wang, Z. *et. al.* Chemical selectivity at grain boundary dislocations in monolayer $Mo_{1-x}W_xS_2$ transition metal dichalcogenides. *ACS Appl. Mater. Interfaces* **9**, 29438–29444 (2017).